\documentclass [a4paper,fleqn, 12pt]{article}
\usepackage{graphicx}
\usepackage[small]{subfigure,epsfig}

\usepackage {amsmath} \usepackage{amssymb} \usepackage{cite}

\newcommand{\cn}

\begin{document}


\title
{A Lagrangian Description of the Higher-Order Painlev\'e Equations}

\author
{A. Ghose Choudhury, \and Partha Guha, \and Nikolai A. Kudryashov}

\date{Department of Physics, Surendranath  College, 24/2 Mahatma
Gandhi Road, Calcutta-700009, India;\\
S.N. Bose National Centre for Basic Sciences,
JD Block, Sector III, Salt Lake, Kolkata - 700098,  India;\\
Department of Applied Mathematics, National Research Nuclear University
MEPHI, 31 Kashirskoe Shosse,
115409 Moscow, Russian Federation}




\maketitle

\begin{abstract}
We derive the Lagrangians of the higher-order Painlev\'e
equations using Jacobi's last multiplier technique. Some of these higher-order
differential equations display certain remarkable properties like passing
the Painlev\'e test and
satisfy the conditions stated by Jur\'a$\check{s}$, (Acta Appl.
Math.  66  (2001) 25--39), thus allowing for a Lagrangian description.

\end{abstract}






\section{Introduction}

The study of higher-order Painlev\'e equations is interesting from the mathematical
point of view because of the possibility of existence of new transcendental functions beyond the six Painlev\'e transcendents.
In addition such higher-order Painlev\'e often have interesting physical and mathematical applications. For example
it is known that special solutions of equations for the Korteweg de Vries hierarchy which are used for describing
water waves can be expressed via the higher-order Painlev\'e equations.

The first Painlev\'e hierarchy was first introduced in \cite{Kud}. Thereafter many results were obtained in the analysis
of the higher Painlev\'e equations. Scaling similarity solutions of three integrable PDEs namely the Sawada-Kotera, fifth
order KdV and Kaup-Kupershmidt equations were considered in \cite{Hone} where it was shown that these fourth-order ordinary differential
equations (ODEs) may be
written as non-autonomous Hamiltonian equations for time dependent generalizations of integrable cases of the H\'enon-Heiles system.

In \cite{Kud1} it was proved that higher-order members for the first and second Painlev\'e hierarchies do not have  polynomial  first
integrals and that their solutions can determine new transcendental functions. Lax pairs for some equations of these hierarchies
are presented in \cite{Kud1a} and the Cauchy problem for equations of these hierarchies can be solved by an analogy with the Cauchy problem
of the well known Painlev\'e equations. The Painlev\'e tests for higher-order Painlev\'e equations were demonstrated in \cite{Mugan01, Kud2, Mugan02}.

In \cite{Kud3} two new hierarchies of nonlinear ODEs were introduced which were called the $K_1$ and $K_2$ hierarchies and which may be
considered as new higher Painlev\'e hierarchies. The equations of these hierarchies have all the properties that are unique to the famous Painlev\'e equations.

Shimomura in \cite{Shimomura01} presented an interesting expression for the first Painlev\'e hierarchy which allows us to consider new properties of equations.
Poles and $\alpha$ - points of the meromorphic solutions of the first Painlev\'e hierarchy was studied by Shimomura in \cite{Shimomura02},
where a lower estimate for the number of poles of meromorphic solution is also given.

In \cite{Aoki} instanton-type solutions and some leading expressions for the
second member of the first hierarchy were constructed using multiple-scale
analysis. Recently Mo in \cite{Mo01} has applied a twistor description of the similarity reductions to the case of the KdV hierarchy
to obtain the twistor spaces of the Painlev\'e I and Painlev\'e II hierarchy.  Dai and Zhang \cite{Dai} have extended
the results by Boutroux \cite{Boutroux1,Boutroux2}
for the first Painlev\'e equation to the case of the first Painlev\'e hierarchy. The authors have shown that there are solutions characterized by
divergent asymptotic expansions near infinity in specified sectors of the complex plane for higher-order analogue of the first Painlev\'e equation.

Some important results connected with higher-order Painlev\'e equations were also obtained in the papers \cite{Claeys1, Claeys2}.
In \cite{Claeys1} Claeys and Vanlessen proved the universality of the correlation kernel in a double scaling limit near singular edge
points in random matrix models that were built out of functions associated with a special solution of the second member
for the first Painlev\'e hierarchy. In \cite{Claeys2} the authors established the existence of real solution of the fourth-order analogue of the Painlev\'e
equation and obtained the solvability of an associated Riemann - Hilbert problem through the approach of a vanishing lemma and found additionally the asymptotics of solutions.

The Hamiltonian structure of the second Painlev\'e hierarchy was considered in \cite{Mo02}. Here the authors introduced new canonical coordinates and obtained the
Hamiltonian for evolutions. They also gave an explicit formulae for these Hamiltonians and demonstrated that these Hamiltonians are polynomials in the canonical coordinates.

\bigskip

The aim of this paper is to obtain the Lagrangians for the four higher Painlev\'e hierarchies using the same approach.
In recent years much attention has been paid to the Lagrangian
framework of higher-order differential equations. Although a
Lagrangian always exists for any second-order ordinary
differential equation its connection with  Jacobi's last
multiplier (JLM) \cite{Jac1, Jac2} is perhaps not very widely
known. The credit for resurrecting the JLM, in recent years, must
go to Leach and Nucci, who have shown how it may be used to
determine the first integrals and also Lagrangians of a wide
variety of nonlinear differential equations \cite{Leach}. While it
appears that the connection of the Jacobi last multiplier to the
existence of Lagrangian functions were the subjects of
investigation by a few authors in the early 1900's, the precise
nature of this interrelation was brought out by Rao, in the 1940's
\cite{Rao}. Thereafter it does not appear to have attracted the
attention of most researchers working in the field of differential
equations.

According to the classical theory of Darboux \cite{Dar} every
scalar second-order ordinary differential equation is
multiplier variational. The problem of finding a Lagrangian  for a
given ODE is generally referred
to as the inverse variational problem of classical mechanics. The
necessary and sufficient conditions for an equation
$y^{\prime\prime}=F(x,y,y^\prime)$ to be derivable from the
Euler-Lagrange equation \label{EL1} $\frac{\partial L }{\partial
y }-\frac{d}{dx}\left(\frac{\partial L}{\partial
y^\prime}\right)=0$, was enunciated by Helmholtz \cite{HH, JL}
in the form of certain identities.

The variational multiplier problem for higher-order scalar
ordinary differential equations has been studied by Fels
\cite{Fels} and Jur\'a$\check{s}$ \cite{Ju}. The inverse problem
for a fourth-order ODE was solved by Fels  who investigated scalar
fourth-order ordinary differential equations of the form
$$
\frac{d^4u}{dx^4} = f(x,u, \frac{du}{dx}, \frac{d^2u}{dx^2},
\frac{d^3u}{dx^3}).
$$

Fels approach for solving the fourth-order inverse problem
was essentially based on a modified version of Douglas's
\cite{Doug} classical solution to the multiplier problem as
refined by Anderson and Thompson in \cite{AT}, who used the
variational bicomplex theory \cite{An} to derive the multiplier
and showed that the existence of a multiplier was in a direct
correspondence with the existence of special cohomology classes
arising in the variational bicomplex associated with a
differential equation. Fels conditions ensure the existence and
uniqueness of the Lagrangian in the case of a fourth-order
equation and it has been shown by Nucci and Arthurs \cite{NA} and
more recently by us \cite{GGK} that when these conditions are
satisfied, a Lagrangian can be derived from the Jacobi last
multiplier.

In fact Fels approached the problem using Cartan's equivalence method, and
arrived at two differential invariants whose vanishing completely
characterizes the existence of a variational multiplier. Unlike
the second-order case, the multiplier is unique up to a constant
multiple. The programme was further developed by Jur\'a$\check{s}$
\cite{Ju} who studied the inverse problem for sixth and
eighth-order equations. In fact  Jur\'a$\check{s}$ obtained a
similar solution by using, however, a more direct approach in the
spirit of the variational bicomplex;  the differential invariants
becoming increasingly complicated for higher-order systems. By
analyzing the structure equations of the horizontal differential
he uncovered a two-form $\Pi$ with the property $d \Pi \equiv 0 \,
\, \mod\, \Pi$, if and only if the equation
$$
\frac{d^{2n}u}{dx^{2n}} = f \big(x,u,\frac{du}{dx}, \cdots,
\frac{d^{2n-1}u}{dx^{2n-1}}\big),
$$
is multiplier variational. He proved that a Lagrangian, if it
exists, is unique up to the multiplication by a constant and found
functions $I_1, I_2, . . . ,I_n$, whose vanishing provides a
necessary condition for the above equation to be variational.
These functions are not relative contact invariants, but their
simultaneous vanishing is a contact invariant condition.

In \cite{GGK} the  authors  made use of the Jacobi Last
 Multiplier (JLM) to derive Lagrangians for a set of fourth-order
 ODEs which pass the Painlev\'{e} test, i.e., their solutions are free of movable
 critical points. Recently the conjugate
Hamiltonian equations for such fourth-order equations passing the
Painlev\'e test have also been  derived in \cite{GGF}.

\section{Four Painlev\'e hierarchies}

Now the first and the second Painlev\'e hierarchy are well known and can be written as the following
\begin{equation}\label{Eq33}
\begin{gathered}
\sum_{m=1}^{N}\,t_{m}\,L_{m}[u]=x,
\end{gathered}
\end{equation}
\begin{equation}\label{Eq44}
\begin{gathered}
\left(\frac{d}{dx}+u\right)\sum_{m=1}^{M}\,t_{m}\,L_{m}[u_x-u^2]-x\,u-\beta_N=0,
\end{gathered}
\end{equation}
where $N$ and $M$ are integers, $t_{m}$, $(m=1,\ldots,N)$ is the sequence of operators $L_{m}[u]$ that satisfies the Lenard recursion relation
\begin{equation}\begin{gathered}
\label{Lenard}d_x\,L_{m+1}[u]=\left(d^{\,3}_{x}+4\,u\,d_x+2\,u_x\right)\,L_m[u],\qquad
L_0[u]=\frac{1}{2}.
\end{gathered}\end{equation}

Taking the operator \eqref{Lenard} into account we obtain
\begin{equation}\begin{gathered}
\label{Lenard_1a}L_{1}[u]=u,
\end{gathered}\end{equation}

\begin{equation}\begin{gathered}
\label{Lenard_2a}L_{2}[u]=u_{xx}+3\,u^2,
\end{gathered}\end{equation}

\begin{equation}\begin{gathered}
\label{Lenard_3a}L_{3}[u]=u_{xxxx}+10\,u\,u_{xx}+5\,u_{x}^2+10\,u^3,
\end{gathered}\end{equation}

\begin{equation}\begin{gathered}
\label{Lenard_4a}L_{4}[u]=u_{xxxxxx}+14\,u\,u_{xxxx}+28\,u_{x}\,u_{xxx}+21\,u_{xx}^2+\\
\\
70\,u^2\,u_{xx}+70\,u\,u_{x}^2+35\,u^4.
\end{gathered}\end{equation}

Using the values of operators $L_1$, $L_2$, $L_3$, $L_4$ and so on we can obtain the equations of the first and the second Painlev\'e hierarchies.

The sixth-order ordinary differential equations of the first and the second Painlev\'e hierarchies have the form

\begin{equation*}\label{Eq1}
\begin{gathered}
{t_4}\, \left( u_{xxxxxx}  +14\,u\,u_{xxxx}
 +28\,u_{x}\,u_{xxx}
 +21\,u_{xx}^{2} +70\,u^2\,u_{xx}+\right.
 \\
 \left.+70\,u\,u_{x}^{2}
  +35\, u^4 \right)
   +\, {t_3}\,
 \left(u_{xxxx}+10\,u \,u_{xx}+5\, u_{x}^2+ 10\,u^3
\right) \,+
\\
+\,{t_2}\, \left( u_{xx}+3\,u^2 \right) +t_1\,u =x,
\end{gathered}
\qquad (A)
\end{equation*}

\begin{equation*}\label{Eq2}
\begin{gathered}
{t_3}\, \left(u_{xxxxxx} -14\,u^2\,u_{xxxx}-56\,u\,u_{x}\,u_{xxx}
 -28\,u_{x}^2\,u_{xx} -42\,u\,u_{xx}^2 +\right.
 \\
 \left.+70\,u^4\,u_{xx}+140\,u^3\,u_{x}^2 -20\,u^7 \right)+
 {t_2}\, \left(u_{xxxx}  -10\, u^2\,u_{xx}-\right.\\
 \left.- 10\,u\,u_{x}^2  +6\,u^5
 \right) + {t_1}\, \left( u_{xx}-2\,u^3\right) -x\,u
 -\beta_3=0,
\end{gathered}\qquad (B)
\end{equation*}

We see that equations of the first and the second hierarchy have even integer orders $2\,N-2$ and $2\,M$ respectively.

Equations (A) and (B) are important and interesting because
setting the constants $t_3=t_2=0$ one recovers the Painlev\'e
equations. When $t_1=t_3=0$ these yield equations which we have
studied recently. In the case $t_1=t_2=0$ they reduce to
sixth-order equations which are the third members of the first and
second Painlev\'{e} hierarchies. The general case of these equations
correspond to the first and second Painlev\'e
hierarchies.

There are two other hierarchies of nonlinear ordinary differential equations that have the properties similar to Painlev\'e equations.
These hierarchies were introduced in \cite{Kud3} and were referred to in \cite{Kud4} as the $K_1$ and $K_2$ hierarchies.
These hierarchies can be presented as the following

\begin{equation}
\begin{gathered}\label{K1}
\sum_{m=1}^{N}\,t_{m}\,H_{m}\left[u\right]=x,
\end{gathered}
\end{equation}

\begin{equation}
\begin{gathered}
\left(\frac{d}{dx}+u\right)\sum_{m=1}^{M}\,t_{m}\,H_{m}\left[u_x-
\frac12\,u^2\right]-x\,u-\beta_M=0,
\label{K2}
\end{gathered}
\end{equation}
where $N$ and $M$ are integers, $t_m$ are parameters of the equation and the operator $H_m$ may be calculated by means of the
formulae
\begin{equation}
\begin{gathered}
H_{n+2}=J[v]\,\Omega[v]\,H_{n}, \label{Eq3}
\end{gathered}
\end{equation}
under the conditions

\begin{equation}
\begin{gathered}\label{Eq3m}
H_{0}[v]=1,\qquad H_{1}[v]=v_{xx}+4\,v^2,
\end{gathered}
\end{equation}
where the operators $\Omega[v]$ and $J[v]$ are  determined by the
relations

\begin{equation}
\begin{gathered}
\Omega=D^3+2\,v\,D+v_x,\qquad D=\frac{d}{dx},
\label{Eq3b}
\end{gathered}
\end{equation}

\begin{equation}
\begin{gathered}
J=D^3+3\,(v\,D+D\,v)+2\,(D^2\,v\,D^{-1}+D^{-1}\,v\,D^2)+\\
\\+8\,(v^2\,D^{-1}+D^{-1}\,v^2),\qquad
D^{-1}=\int\,dx
\label{Eq3c}.
\end{gathered}
\end{equation}

Taking conditions \eqref{Eq3m} and operators \eqref{Eq3b}, \eqref{Eq3c} into account we have the operators $H_2$ and $H_3$ as the following

\begin{equation}
\begin{gathered}
H_2[v]=v_{xxxx}+12\,v\,v_{xx}+6\,v_{x}^2+\frac{32}{3}\,v^3,
\label{Eq3d}
\end{gathered}
\end{equation}

\begin{equation}
\begin{gathered}
H_3[v]=v_{xxxxxxxx}+20\,v\,v_{xxxxxx}+60\,v_{x}\,v_{xxxxx}+134\,v_{xx}\,v_{xxxx}+\\
\\+
136\,v^2\,v_{xxxx}
+84\,v_{xxx}^2+544\,v\,v_{x}\,v_{xxx}+408\,v\,v_{xx}^2+396\,v_{x}^2\,v_{xx}+\\
\\+
\frac{1120}{3}v^3\,v_{xx}+560\,v^2\,v_{x}^2+\frac{256}{3}\,v^5.
\label{Eq3e}
\end{gathered}
\end{equation}

Note that hierarchies \eqref{K1} and \eqref{K2} can also be presented using another operator $G_{k}[u]$. In terms of this operator these hierarchies take in the form

\begin{equation}
\begin{gathered}
\sum_{k=1}^{N}\,t_{k}\,G_{k}[u]=x.
\label{K1a}
\end{gathered}
\end{equation}

\begin{equation}
\begin{gathered}
\left(u-\frac12\,\frac{d}{dx}\right)\sum_{k=1}^{M}\,t_{k}\,G_{k}[-2\,u_x-
2\,u^2]-x\,u-\beta_M=0.
\label{K2a}
\end{gathered}
\end{equation}

Hierarchy \eqref{K1a} can be transformed to \eqref{K1} but hierarchy \eqref{K2a} coincides with hierarchy \eqref{K2}.
The recursion relation $G_k$ is determined  by the nonlinear operator
\begin{equation}
\begin{gathered}
G_{k+2}=J_1[v]\,\Omega[v]\,G_k, \label{Eq6}
\end{gathered}
\end{equation}
under the conditions
\begin{equation}
\begin{gathered}
G_0[v]=1,\qquad G_1[v]=v_{xx}+\frac14\,v^2. \label{Eq6}
\end{gathered}
\end{equation}
The operator $J_1[v]$ takes the form
\begin{equation}
\begin{gathered}
J_1=D^3+\frac12\,(D^2\,v\,D^{-1}+D^{-1}\,v\,D^2)+\frac18\,(v^2\,D^{-1}+D^{-1}\,v^2).
\label{Eq7}
\end{gathered}
\end{equation}

Hierarchies $K_1$ and $K_2$ though similar to the first and the second Painlev\'e hierarchies have a fundamental difference in the sense that
 we cannot transform equations of hierarchies \eqref{K1a} and \eqref{K2a} to hierarchies \eqref{Eq33} and \eqref{Eq44}.
Moreover the hierarchy $K_1$ has even integer order except $6\,k\,\,\,(k=1,2,\ldots)$ and hierarchy $K_2$ also has even integer order except $6\,k\,\,\,(k=1,2,\ldots)$.

The fourth order equation corresponding to the hierarchy $K_1$ takes the form

\begin{equation*}\label{Eq3a}
\begin{gathered}
{t_2}\,\left(u_{xxxx}+12\,u\,u_{xx}+6\,u_{x}^2+\frac{32}{3}\right)
 +{t_1}\,\left(u_{xx}+4\,u^2\right) =0.
\end{gathered} \qquad(C)
\end{equation*}

At $t_2=0$ equation (C) is the first Painlev\'e equation but at $t\neq 0$ the forth order equation differs from the the fourth order equation of the first Painlev\'e equation and we hope that this one may give a new transcendal function.

On the other hand the sixth-order equation from hierarchy $K_2$ can be written as
\begin{equation*}\label{Eq3}
\begin{gathered}
{t_2}\, \left(u_{xxxxxx}+7\,u_x\,u_{xxxx}-7\,u^2\,u_{xxxx}+ 14\,u_{xx}\,u_{xxx}
-28\,u\,u_x\,u_{xxx}-\right.\\
\left. -28\,u_{x}^2\,u_{xx} -21
\,u\,u_{xx}^2 -{\frac {28}{3}}\,u\,u_{x}^3 -14\,u^2\,u_x\,u_{xx}+14\,u^4\,u_{xx}
 +\right.\\
 \left.+28\,u^3\,u_{x}^2 -\frac43\, u^7 \right)
 +{t_1}\, \left( u_{xxxx} +5\,u_x\,u_{xx} -5\,u^2\,u_{xx} -\right.\\\left.-5\,
u\,u_{x}^2 + u^5
 \right) -x\,u  -\beta_2=0.
\end{gathered}\qquad (D)
\end{equation*}

Equation (D) is a sixth-order nonlinear ordinary differential
equation with  properties similar to the Painlev\'e equations  but cannot be
transformed to the equation of the second Painlev\'e hierarchy.
This equation does not have a  first integral in the polynomial
form and it is possible  that it determines
a new transcendental function.

\smallskip

In the following section we find the Lagrangians for the nonlinear ordinary differential equations (A), (B) and (D).

\section{Inverse problem for sixth-order equations and their Lagrangians}

Consider a sixth-order equation in the normal form, $ u_6 =
f(x,u,u_1,u_2,u_3,u_4,u_5)$.  Here we introduce the abridged
notation $u_k=d^ku/dx^k$. The following theorem due to
Jur\'a$\check{s}$ gives the necessary and sufficient conditions
for a sixth-order equation to admit a variational multiplier
\cite{Ju}.

\textbf{Theorem.}
A sixth-order ordinary differential equation admits a variational
multiplier and non-degenerate third-order Lagrangian if and only
if following two conditions are satisfied
$$
 0 = -\frac{2}{3}D_{x}^{4}\big(\frac{\partial f}{\partial u_5} \big) + \frac{10}{9}\frac{\partial f}{\partial u_5}
D_{x}^{3}\big(\frac{\partial f}{\partial u_5} \big) + D_{x}^{3}\big(\frac{\partial f}{\partial u_4} \big)
+ \frac{20}{9} D_x\big(\frac{\partial f}{\partial u_5} \big) D_{x}^{2}\big(\frac{\partial f}{\partial u_5} \big)
$$
$$
- \frac{20}{27}\big(\frac{\partial f}{\partial u_5} \big)^2 D_{x}^{2}\big(\frac{\partial f}{\partial u_5} \big)
- \frac{1}{3}\frac{\partial f}{\partial u_4} D_{x}^{2}\big(\frac{\partial f}{\partial u_5} \big) -
\frac{\partial f}{\partial u_5}
D_{x}^{2}\big(\frac{\partial f}{\partial u_4} \big) - D_{x}^{2}\big(\frac{\partial f}{\partial u_3} \big)
$$
$$
- \frac{10}{9}\frac{\partial f}{\partial u_5}
\big(D_x\big(\frac{\partial f}{\partial u_5} \big)\big)^2 -
D_x\big(\frac{\partial f}{\partial u_5}
\big)D_x\big(\frac{\partial f}{\partial u_4} \big) +
\frac{20}{81}\big(\frac{\partial f}{\partial u_5} \big)^3
D_x\big(\frac{\partial f}{\partial u_5} \big)
$$
$$
+ \frac{1}{3}
\big(\frac{\partial f}{\partial u_5} \big)^2 D_x\big(\frac{\partial f}{\partial u_4} \big) +
\frac{1}{3}\frac{\partial f}{\partial u_5}
\frac{\partial f}{\partial u_4}D_x \big(\frac{\partial f}{\partial u_5} \big) +
\frac{1}{3}\frac{\partial f}{\partial u_3}
D_x\big(\frac{\partial f}{\partial u_5} \big) +
\frac{2}{3}\frac{\partial f}{\partial u_5}D_x \big(\frac{\partial f}{\partial u_3} \big)
$$
$$
+ D_x\big(\frac{\partial f}{\partial u_2} \big) -
\frac{2}{243}\big(\frac{\partial f}{\partial u_5} \big)^5 -
\frac{1}{27} \big(\frac{\partial f}{\partial u_5}
\big)^3\frac{\partial f}{\partial u_4} -
\frac{1}{9}\big(\frac{\partial f}{\partial u_5} \big)^2
\frac{\partial f}{\partial u_3} - \frac{1}{3}\frac{\partial
f}{\partial u_5}\frac{\partial f}{\partial u_2} - \frac{\partial
f}{\partial u_1},
$$

and
$$
0 = \frac{5}{3}D_{x}^{2}\big(\frac{\partial f}{\partial u_5}\big) - \frac{5}{3}\frac{\partial f}{\partial u_5}
D_x\big(\frac{\partial f}{\partial u_5} \big) -
2D_x\big(\frac{\partial f}{\partial u_4} \big) +
$$
$$
\frac{5}{27} \big(\frac{\partial f}{\partial u_5} \big)^3 +
\frac{2}{3}\frac{\partial f}{\partial u_5}\frac{\partial
f}{\partial u_4} + \frac{\partial f}{\partial u_3}.
$$

\textbf{Prove.}
Suppose the sixth-order equation
$$u_6 = f(x,u,u_1,u_2,u_3,u_4)$$ is independent of $u_5$. Then it
admits a variational multiplier and a non-degenerate third-order
Lagrangian if and only if the following two conditions are satisfied:\\
$ 0 = D_{x}^{3}\big(\frac{\partial f}{\partial u_4} \big) -
D_{x}^{2}\big(\frac{\partial f}{\partial u_3}\big) +
D_x\big(\frac{\partial f}{\partial u_2} \big) - \frac{\partial
f}{\partial u_1}$  and $ 0 = -2D_x\big(\frac{\partial
f}{\partial u_4}\big) + \frac{\partial f}{\partial u_3}.$

\subsection{The Jacobi Last Multiplier and construction of Lagrangians for sixth-order
equations}

In this section we describe the connection of the Jacobi Last
Multiplier with the Lagrangian function for sixth-order ODEs.

\textbf{Proposition.}{}
 Let  $u_6 = f(x,u,u_1,u_2,u_3, u_4, u_5)$ be a
sixth-order ordinary differential equation which admits a
Lagrangian $L$. Then the function $\label{JLM2} {\cal
M}:=\left(\frac{\partial^2 L}{\partial u_{3}^{2}}\right)^3,$ is
a Jacobi last multiplier, i.e., it satisfies the equation
$\label{JLM1}\frac{d{\cal M}}{dx}+\frac{\partial f}{\partial
u_5}{\cal M}=0,$ where $u_5=u_{xxxxx}$.

{\bf Proof :} Considering the higher-order Euler operator, $E$,
the Euler-Lagrange equation of motion for the ODE
$u_6=f(x,u,u_1,...,u_5)$ is given by
\begin{equation}\label{EC}
E(L)
= \frac{\partial L}{\partial u} - D_x\big(\frac{\partial L
}{\partial u_1}\big) + D_{x}^{2}\big(\frac{\partial L}{\partial
u_2}\big) - D_{x}^{3}\big(\frac{\partial L}{\partial
u_3}\big)=0,
\end{equation}
where $L=L(x,u,u_1,u_2,u_3)$ is a third-order
Lagrangian. It is obvious that the partial derivatives of $L$,
namely $L_u, L_{u_1},....L_{u_3}$ are all functions of
$x,u,...,u_3$. Upon expanding the Euler-Lagrange equation we find
that
$$0=E(L)=u_5L_{u_2u_3}-[2u_5L_{u_3u_3x}+u_5L_{u_2u_3}+f(x,u,u_1,...,u_5)
L_{u_3u_3}+u_5D_x(L_{u_3u_3})+$$
$$
+2u_4u_5L_{u_3u_3u_3}+2u_5u_1L_{u_3u_3u}+2u_2u_5L_{u_3u_3u_1}+2u_3u_5L_{u_3u_3u_2}]+
\mbox{terms
independent of }\; u_5.
$$
Here the subscripts denote partial derivatives with respect to the
indicated variables. Since the partial derivative of this equation
with respect to $u_5$ must also be identically zero, we find that
$ \label{jlm3} 3D_x(L_{u_3u_3})+\frac{\partial f}{\partial
u_5}(L_{u_3u_3})=0.$

Let be  $M_{(3)}=L_{u_3u_3}$,  then the above
equation, $E(L) = 0$ is expressible as be $D_x\left(\log
M_{(3)}^3\right)+\frac{\partial f}{\partial u_5}=0$, showing
thereby that the Jacobi Last multiplier is given by
$${\cal M}=M_{(3)}^3.\;\;\;\;\;\Box$$

\smallskip

\textbf{Remark:} Note that for the fourth-order ODE,
$u_4=f(x,u,...,u_3)$, admitting a second-order Lagrangian the
analog of (\ref{jlm3}) is the following equation:
$$D_x\left(\log
M_{(2)}^2\right)+\frac{\partial f}{\partial u_3}=0, $$  so that
the JLM is ${\cal M}=M_{(2)}^2$ where $M_{(2)}=L_{u_2u_2}$. On the
other hand for the second-order ODE, $u_2=f(x,u,u_1)$, it is the
solution of
$$D_x\left(\log
M_{(1)}\right)+\frac{\partial f}{\partial u_1}=0, $$ with ${\cal
M}=M_{(1)}=L_{u_1u_1}$.

Equation (\ref{EA})  provides us a tool for determining the
Lagrangian of a fourth-order equation once a solution of the
defining equation for the JLM , ${\cal M}$, is obtained from
(\ref{EB}). In fact in the event $f$ is independent
of $u_5$, so that the condition (\ref{EB}) is trivially
satisfied one obtains the solution $M_{(3)}=$ constant, which may
be set equal to unity, without loss of generality. In such a
situation the Jur\'a$\check{s}$ conditions are also considerably
simplified as evident from the Corollary 1.

\subsection{Determination of the Lagrangians}

We wish to determine a nondegenerate third-order Lagrangian $L =
L(x,u,u_1,u_2,u_3)$ such that $\label{EL1} E(L) = 0, \, \, \,
\hbox{ where } \, \, \, \frac{\partial^2 L}{\partial {u_3}^2} \neq
0, $ where $E$ is the Euler-Lagrange operator $\label{EL.1} E
= \frac{\partial}{\partial u} - D_x\big(\frac{\partial}{\partial
u_1}\big) + D_{x}^{2}\big(\frac{\partial}{\partial u_2}\big) -
D_{x}^{3}\big(\frac{\partial}{\partial u_3}\big),$ and $D_x$
denotes the total derivative operator $ D_x =
\frac{\partial}{\partial x} + u_1\frac{\partial}{\partial u} +
u_2\frac{\partial}{\partial u_1} + u_3\frac{\partial}{\partial
u_2}.$ If there is a third-order Lagrangian satisfying
the conditions stated in Theorem 3.1, one says that the ordinary differential
equation $u_6 = f(x,u,u_1,u_2,u_3,u_4,u_5)$ admits a variational
multiplier.

In the new notation equation (A) is given by
$$t_3(u_6+14uu_4+28u_1u_3+21u_2^2+70u^2u_2+70uu_1^2+35u^4)+$$
$\label{6.1} +t_2(u_4+10uu_2+5u_1^2+10u^3)+t_1(u_2+3u^2)=x.$

\textbf{Proposition.}{}
Equation (A) admits a Lagrangian description with Lagrangian
$\label{6Lag}L=t_3(-\frac{1}{2}u_3^2+7u^5-35u^2u_1^2+7uu_2^2)
+t_2(\frac{1}{2}u_2^2-5uu_1^2+\frac{5}{2}u^4)+t_1(-\frac{1}{2}u_1^2+u^3)-x\;u,$
where $u_k=u^{(k)},\;\; k=1,2,....$.

{\bf Outline of the proof :} In order to show this we will adopt the technique used in
 \cite{GGK}, to derive Lagrangians for a certain class of
 fourth-order ODEs, namely that of the Jacobi Last Multiplier
 (JLM). For a sixth-order ODE written in the form
 $$u_6=f(x,u,u_1,...,u_5)$$ one can rewrite the equation as a
 first-order system:
$\label{jlm.1}u_1=v,\;v_1=w,\;w_1=s,\;s_1=t,\;t_1=r,\;
r_1=f(x,u,v,w,s,t,r),$ where the subscript $1$ denotes
differentiation with respect to the independent variable $x$. Then
by definition the JLM, ${\cal M}$, for the above system of
first-order ODEs is defined as the solution of the following
equation $\label{jlm.2}\frac{d\log {\cal
M}}{dx}+\left(\frac{\partial v}{\partial u }+\frac{\partial
w}{\partial v }+\frac{\partial s}{\partial w }+\frac{\partial
t}{\partial s }+\frac{\partial r}{\partial t }+\frac{\partial
f}{\partial r }\right)=0.$ Since in case of eqns (A) and (B) the
function $f$ is independent of $u_5$ i.e., of $r$ in this notation
it follows that $\label{jlm.3}\frac{d\log
M_{(3)}}{dx}=0\;\;\Rightarrow\;\;M_{(3)}=constant.$ Furthermore
since the JLM is connected to the Lagrangian by the
following relation  $\label{jlm.4}M_{(3)}=\frac{\partial^2
L}{\partial u_3^2},\;\;\mbox{where}\;\;u_3=\frac{d^3u}{dx^3},$
therefore setting the constant in (\ref{EA}) to be $-t_3$ we
have
$$\frac{\partial^2
L}{\partial u_3^2}=-t_3\;\Rightarrow
L=-t_3\frac{u_3^2}{2}+R(x,u,u_1,u_2) u_3+V(x,u,u_1,u_2).$$

Consequently
the determination of the Lagrangian essentially reduces to finding
appropriate functions $R$ and $V$ such that the Euler-Lagrange
equation $E(L)=0$, reproduces
the desired equation, namely (\ref{EA}). Detailed calculations
however show that it is possible to choose $R=0$ so as to simplify
the resulting calculations, and therefore the Lagrangian is of the
form,
$\label{LagA.1}L_A=t_3(-\frac{1}{2}u_3^2)+V(x,u,u_1,u_2).$ To
deduce the unknown function $V$ we substitute this form of the
Lagrangian into (\ref{EC}) and compare the resulting equation
with our original sixth-order ODE. One finds that in this case
$$E(L_A)=\frac{\partial L_A}{\partial u}-D_x\left(\frac{\partial L_A}{\partial u_1}\right)
+D_x^2\left(\frac{\partial L_A}{\partial
u_2}\right)-D_x^3\left(\frac{\partial L_A}{\partial
u_3}\right)=0$$ gives
$$-t_3u_6=V_u-V_{xu_1}-u_1V_{uu_1}-u_2V_{u_1u_1}+u_4V_{u_2u_2}+V_{xxu_2}
+2u_1V_{xuu_2}+2u_2V_{xu_1u_2}+2u_3V_{xu_2u_2}$$
$$+u_1^2V_{uuu_2}+u_2^2V_{u_1u_1u_2}+u_3^2V_{u_2u_2u_2}
+2u_1u_2V_{uu_1u_2}+2u_1u_3V_{uu_2u_2}+2u_2u_3V_{u_1u_2u_2}.$$
Inserting the value of $u_6$ from the original equation and
equating the coefficient of $u_4$ we find that
$$V_{u_2u_2}=t_314u+t_2$$ which leads to the following solution,
namely $\label{LagA.2}V=t_37uu_2^2+t_2\frac{u_2^2}{2}+Tu_2+S$
where $T$ and $S$ are functions of $x,u,u_1$. Once again we may
 set $T=0$ so that
$\label{lagA.3}L_A=t_3\left(-\frac{1}{2}u_3^2+7uu_2^2\right)+t_2(\frac{1}{2}u_2^2)+S,$
and it remains therefore to determine the unknown function $S$.
From the remaining terms we find that one must have
$\label{LagA.4}S_u-S_{xu_1}-u_1S_{uu_1}-u_2S_{u_1u_1}=t_3(70u^2u_2+70uu_1^2+35u^4)
+t_2(10uu_2+5u_1^2+10u^3)+t_1(u_2+3u^2)-x.$ Next equating the
coefficients of $u_2$ it is seen that $\label{LagA.5}-S=t_3
35u^2u_1^2+t_25uu_1^2+t_1\frac{u_1^2}{2}+Ku_1+N(x,u).$ Again
choosing  $K=0$, we have ultimately from (\ref{LagA.4}),
$$-N_u=t_335u^4+t_2 10u^3+t_13u^2-x.$$ This yields
$$-N=t_37u^5+t_2 \frac{5}{2}u^4+t_1u^3-xu,$$ and finally the
following expression for the unknown function $S$
$$S=-t_335u^2u_1^2-t_25uu_1^2+t_1(-\frac{1}{2}u_1^2)+t_37u^5+t_2\frac{5}{2}u^4+
t_1u^3-xu.$$
We find finally that the
 expression for the Lagrangian of eqn. (A) is,
$$L_A=t_3\left(-\frac{1}{2}u_3^2+7u^5-35u^2u_1^2+7uu_2^2\right)
+t_2\left(\frac{1}{2}u_2^2-5uu_1^2+\frac{5}{2}u^4\right)+t_1\left(-
\frac{1}{2}u_1^2+u^3\right)-x\,u.$$
$\Box$

\bigskip

In a similar manner we find that the sixth-order equations (B) and (D)
also admit a Lagrangian description, which are stated below.

\textbf{Proposition.}
The Lagrangians associated with the equations (B) and (D) are given by
\begin{equation}
\begin{gathered}
L_B=t_1\left(-\frac{1}{2}u_1^2-\frac{1}{2}u^4\right)+
t_2\left(\frac{1}{2}u_2^2+5u^2u_1^2+u^6\right)+\\
+t_3\left(-\frac{1}{2}u_3^2-\frac{5}{2}u^8-35u^4u_1^2-
7u^2u_2^2\right)-\frac12\,x\,u^2-\beta_3
u,
\end{gathered}
\end{equation}
$$L_D=t_2\left(-\frac{1}{2}u_3^2+\frac{7}{2}u_1u_2^2-\frac{7}{2}u^2u_2^2+
\frac{7}{6}u_1^4+\frac{7}{3}u^2u_1^3-7u^4u_1^2-\frac{u^8}{6}\right)+$$
\begin{equation}+t_1\left(\frac{1}{2}u_2^2+\frac{5}{2}u^2u_1^2-
\frac{5}{6}u_1^3+\frac{u^6}{6}\right)-
\frac{1}{2}\,x\,
u^2-\beta_2u.
\end{equation}

\bigskip

\section{Conclusion}

Let us briefly discuss the results of this paper. We have found
the Lagrangians $L_A$, $L_B$  and $L_D$ for the sixth-order
nonlinear differential equations from the Painlev\'e hierarchies.
These Lagrangians are generalizations of the well known
Lagrangians of the Painlev\'e equations. It is interesting to look
at the properties of the Lagrangians $L_A$, $L_B$ and $L_D$. For
 example we know that there exists the following  symmetry of equation (B) when $u(x)
\rightarrow -u(x)$ and $\beta_3 \rightarrow -\beta_3$. This
symmetry is preserved for the Lagrangian $L_B$ as well. Equations
(A), (B) and (D) posses the Painlev\'e property. The Lagrangians
of several mechanical systems usually involve a   difference of
their kinetic and potential energies respectively.
 Here this property holds for the
Lagrangians $L_A$ and $L_B$ at $t_2=t_3=x=0$. However  this is not
true in the general case and as such the Lagrangians derived here
are really examples of nonstandard ones. The question of irreducibility of the higher
Painlev\'e equations is an open problem.

\section*{Acknowledgement}
We wish to thank Pepin Cari\~nena, Thanasis Fokas, Peter Leach and Mark Fels for their valuable comments.
  One of us (AGC) wishes to acknowledge the support
provided by the S. N. Bose National
 Centre for Basic Sciences, Kolkata in the form of an
 Associateship.

\end{document}